\documentclass[aps,prb,twocolumn,showpacs,preprintnumbers,amsmath,amssymb,superscriptaddress]{revtex4}
\usepackage{graphicx}
\usepackage{tabularx}
\usepackage{color}
\usepackage{xspace}

\newcommand{\txtr}{\textcolor{red}}

\newcommand{\YBCOx}{YBa$_2$Cu$_3$O$_x$\xspace}

\begin{document}
\preprint{PREPRINT (\today)}

\title{ Muon spin rotation investigation of the pressure effect on the magnetic penetration depth in YBa$_2$Cu$_3$O$_x$  }

\author{A.~Maisuradze}\email{alexander.maisuradze@psi.ch}
\affiliation{Laboratory for Muon Spin Spectroscopy, Paul Scherrer Institut, CH-5232 Villigen PSI, Switzerland}
\affiliation{Physik-Institut der Universit\"{a}t Z\"{u}rich, Winterthurerstrasse 190, CH-8057 Z\"{u}rich, Switzerland}
\author{A.~Shengelaya}
\affiliation{Department of Physics, Tbilisi State University, Chavchavadze av. 3, GE-0128 Tbilisi, Georgia}
\author{A. Amato}
\affiliation{Laboratory for Muon Spin Spectroscopy, Paul Scherrer Institut, CH-5232 Villigen PSI, Switzerland}
\author{E.~Pomjakushina}
\affiliation{Laboratory for Developments and Methods, Paul Scherrer Institut, CH-5232 Villigen PSI, Switzerland}
%
%
\author{H.~Keller}
\affiliation{Physik-Institut der Universit\"{a}t Z\"{u}rich, Winterthurerstrasse 190, CH-8057 Z\"{u}rich, Switzerland}

\begin{abstract}
The pressure dependence of the magnetic penetration depth $\lambda$ in
polycrystalline samples of YBa$_{2}$Cu$_{3}$O$_x$
with different oxygen concentrations $x = 6.45$, 6.6, 6.8, and 6.98 was studied by
muon spin rotation ($\mu$SR).
The pressure dependence of the superfluid density $\rho_s \propto 1/\lambda^2$ as a function of the
superconducting transition temperature $T_{\rm c}$ is found to deviate from the usual Uemura line. The
ratio $(\partial T_{\rm c}/\partial P)/(\partial \rho_s/\partial P)$ is factor of $\simeq 2$
smaller than that of the Uemura relation.
In underdoped samples, the zero temperature superconducting gap $\Delta_0$ and the BCS
ratio $\Delta_0/k_B T_{\rm c}$
both increase with increasing external hydrostatic pressure,
implying an increase of the coupling strength with pressure.
The relation between the pressure effect and the oxygen isotope effect on $\lambda$
is also discussed. In order to analyze reliably the $\mu$SR spectra of samples with strong
magnetic moments in a pressure cell, a special model was developed and applied.
\end{abstract}

\maketitle

\section{Introduction}
The compound YBa$_{2}$Cu$_{3}$O$_x$ was the first high temperature superconductor\cite{Bednorz86} (HTS)
with a superconducting transition temperature $T_{\rm c}$ above the boiling point of
liquid nitrogen, and is one of the most studied HTSs.\cite{Chu87}
Its superconducting properties are well characterized, even though  some
of them are still being heavily discussed. Detailed muon spin rotation
($\mu$SR) studies of the magnetic penetration depth $\lambda$ and
the superfluid density $\rho_s \propto 1/\lambda^2$
were performed on poly- and single crystals of YBa$_{2}$Cu$_{3}$O$_x$ at ambient
pressure.\cite{Harschman87, Brewer88, Uemura88, Uemura89, Pumpin90, Zimmermann95, Riseman95, Tallon95}
However, the key question concerning the pairing mechanism responsible
for high temperature superconductivity is still not resolved, and is subject
of intense debates. Although it is widely believed that magnetic fluctuations play a dominant role
in the pairing mechanism,\cite{Lee06_RMP} oxygen isotope effect (OIE) studies indicate
that lattice degrees of freedom  are essential  for the occurence of
superdonductivity.\cite{Franck94, Zhao98_OIE, Temprano00, Zhao01_OIE, Khasanov03_OIE_JPCM, Keller05, Khasanov08_OIE_gap, Khasanov08_PhDiagr, Keller08_MT}
By means of isotope substitution one can probe the
influence of lattice degrees of freedom on superconductivity without changing
the lattice parameters.\cite{Mali02}
There are no other easily accessible methods which allow to solely modify the exchange integral $J$,
in order to investigate its influence on the superconducting state.\cite{Andre10}
However, the application of hydrostatic pressure changes the interatomic distances in the lattice
which in turn modifies both the lattice dynamics\cite{Calamiotou09} and the exchange coupling $J$
between the Cu spins in cuprates.\cite{Harrison80,Amato08}
Therefore, a detailed study of the pressure effect (PE) on the superconducting
properties, {\it e.g.,} the superfluid density $\rho_s \propto 1/\lambda^2$, the gap magnitude $\Delta_0$,
and the  BCS ratio $\Delta_0/k_BT_{\rm c}$, may provide important information for testing
microscopic theories of the high-temperature superconductivity.\cite{Schilling92,Takahashi96}

Up to now, the PE on the superconducting transition temperature $T_{\rm c}$
was studied by resistivity and Hall effect
experiments.\cite{Almasan92,Rusiecki90,Parker88,Murayama91}
Several phenomenological\cite{Almasan92,Gupta95,Neumeier93} and microscopic models were
proposed based on a Hubbard\cite{Angilella96,Mello97}
or a general BCS approach in order to explain the PE on $T_{\rm c}$.\cite{Chen00}
The role of nonadiabatic effects is discussed in Ref. \onlinecite{Sarkar98}.
These models suggest two basic sources for the PE on $T_{\rm c}$:
(i) A charge transfer from the charge reservoir to the superconducting CuO$_2$ plane,
which was confirmed by Hall effect experiments,\cite{Parker88,Murayama91} and
(ii) an increase of $T_{\rm c}$ due to a pressure dependent pairing interaction.


The magnetic penetration depth $\lambda$ is a fundamental parameter of a superconductor.
It is a measure of the superfluid density according to the relation $1/\lambda^2\propto n_s/m^*$,
where $n_s$ is the superconducting carrier dansity and $m^*$ is the corresponding effective mass.\cite{Uemura88}
From the temperature or field dependence of $\lambda$ one
can determine the symmetry of the superconducting gap, its magnitude and the BCS ratio.
The pressure dependence of $\lambda$ was previously studied in fine powdered grains of
YBa$_{2}$Cu$_{3}$O$_x$\cite{DiCastro09} and YBa$_2$Cu$_4$O$_8$
\cite{Khasanov05_P124,Khasanov04_P124b,Khasanov04_P124c} by means of magnetization experiments.
The $\mu$SR technique is powerful and direct method to determine $\lambda$
in the bulk of a type-II superconductor.\cite{Blundell99muSR,Sonier00RMP}
However, due to several technical difficulties only a small amount of $\mu$SR
studies of the penetration depth under pressure were performed so far.
The main technical problems are: (i) The low fraction of muons stopping
in the sample inside the pressure cell and (ii) the strong diamagnetism of a superconductor
which substantially influences the $\mu$SR response of the pressure cell.

Here, we report on pressure dependent magnetic penetration depth  studies in
polycrystalline samples of YBa$_{2}$Cu$_{3}$O$_x$ ($x = 6.45$, 6.6, 6.8, and 6.98)
by means of $\mu$SR.
We found that the pressure-dependent superfluid density
$\rho_s \propto 1/\lambda^2$ vs $T_{\rm c}$  does not follow the Uemura relation.\cite{Uemura89}
The ratio $\alpha_{\rm p} = (\partial T_{\rm c}/\partial P)/(\partial \rho_s/\partial P)$
is a factor $\simeq 2$ smaller than that of the Uemura relation, but is quite close  to that found
in oxygen isotope effect (OIE) studies,\cite{Khasanov03_OIE_JPCM, Keller05} suggesting a
strong influence of pressure on the lattice degrees of freedom.
Interestingly, a small pressure dependence of the superluid density was also found in the overdoped sample ($x=6.98$).
The superconducting gap $\Delta_0$ and the BCS ratio $\Delta_0/k_BT_{\rm c}$
both increase upon increasing the hydrostatic
pressure in the underdoped samples, hence implying an increase of the coupling strength with pressure.
Finally, a method of data analysis for tranverse-field $\mu$SR measurements of magnetic/diamagnetic
samples loaded in a pressure cell is presented and applied here. This method leads to a substantial
reduction of systematic errors in the data analysis.

The paper is organized as follows: In Sec. II we give some experimental details. In Sec.
III we describe the method of $\mu$SR data analysis and present the experimental results, followed
by a discussion in Sec. IV.
The conclusions are given in Sec. V. In the Appendix we describe the method
used in this work in order to analyze $\mu$SR spectra obtained for a magnetic/superconducting
sample loaded in a pressure cell.

\section{Experimental details}

High quality polycrystalline YBa$_{2}$Cu$_{3}$O$_x$ samples with $x = 6.98$, 6.8, 6.6, and 6.45
were prepared from the starting oxides and carbonate Y$_2$O$_3$, CuO
and BaCO$_3$ as described elsewhere.\cite{Maisuradze09EPR}
Transverse field (TF) $\mu$SR experiments were performed at the $\mu$E1 and $\pi$M3 beam lines of
the Paul Scherrer Institute (Villigen, Switzerland). The samples
were cooled in TF down to 3\,K, and $\mu$SR spectra were taken with increasing temperature
in applied fields $B_\mathrm{app} = 0.1$ and 0.5~T.
Typical statistics for a $\mu$SR spectrum were $5-6\times 10^6$ positron events in the forward and the backward
histograms.\cite{Blundell99muSR, Sonier00RMP}
A CuBe piston-cylinder pressure cell was used with Dafne oil as a pressure transmitting medium.
The maximum pressure achieved was 1.4~GPa at 3~K. The pressure was
measured by tracking the superconducting transition of a very small indium plate used as a manometer
(calibration constant for In: $\partial T_{\rm c}/\partial P = -0.364$~K/GPa).
In order to avoid charge transfer effects due to chain reordering in pressurized
YBa$_{2}$Cu$_{3}$O$_x$, the samples were cooled down
below 100~K for the $\mu$SR measurements within less than 1 hour
after application of the pressure. This time is much shorter than the time constant $\tau = 27.7$~h
(at room temperature) for the pressure activated chain reordering process.\cite{Fita02} Below 100~K
$\tau$ is much longer than the typical measurement time of a sample ($< 24$~h).\cite{Fita02}

High energy muons ($p_\mu \simeq 100$~MeV/c) were implanted in the sample.
Forward and backward positron detectors with respect to the initial muon polarization
were used for the measurements of the $\mu$SR asymmetry time spectrum $A(t)$
(see Fig.~\ref{fig:SampleInPcell}).\cite{Blundell99muSR} Cylindrically pressed
samples were loaded into the cylindrical CuBe pressure cell. The sample dimensions
(diameter 5~mm, height 15~mm) were chosen to maximize the filling factor of the pressure cell.
The fraction of the muons stopping in the sample was approximately 40\%.

\section{Results and analysis details}

For type-II superconductors in the vortex state in an applied field of $B_\mathrm{app}\ll B_{c2}$
($B_{c2}$ is the upper critical field) the square root of the second moment of the
muon depolarization rate $\sigma$ is inversely proportional to the square of the magnetic
penetration depth: $\sigma \propto 1/\lambda^2$ (Refs. \onlinecite{Brewer88,Brandt88,Brandt03})
and therefore directly related to the
superfluid density: $\rho_s \propto 1/\lambda^2 \propto \sigma$.
For a polycrystalline sample of a highly anisotropic and uniaxial superconductor the dominant
contribution to the muon depolarization originates from the in-plane magnetic penetration depth
$\lambda_{\rm ab} = \lambda_{\rm eff}/1.31$, where $\lambda_{\rm eff}$ is an effective (averaged) magnetic
penetration depth.\cite{Barford88, Fesenko91}

\begin{figure}[!tb]
\includegraphics[width=0.95\linewidth]{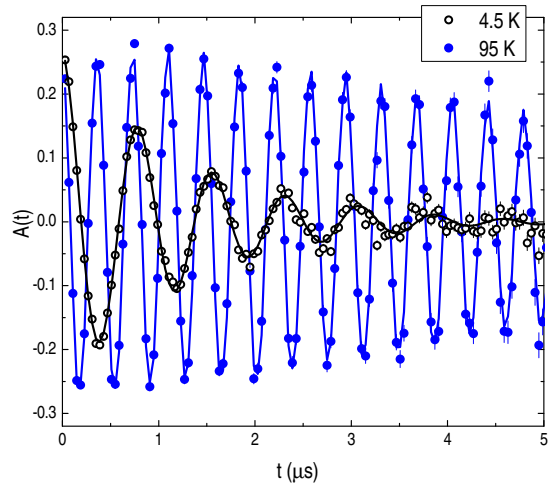}
\caption{(Color online) $\mu$SR asymmetry signal $A(t)$ of YBa$_2$Cu$_3$O$_{6.98}$ measured at
$T = 4.5$~K and 95~K in an applied field $B_\mathrm{app} = 0.1$~T (empty and full circles, respectively).
The fast relaxation of the $\mu$SR signal (empty circles) is due to the formation of a
vortex lattice in the superconducting state. The solid lines are fits of the data to Eq. (\ref{eq:AnFunc}).
For a better visualization the spectra and the fits are shown in a rotating reference frame of 0.08~T.
\label{fig:Fig1asy}}
\end{figure}

As was pointed above a substantial fraction of the $\mu$SR asymmetry signal originates
from muons stopping in the CuBe material surrounding the sample. The sample in the superconducting
state induces an inhomogeneous field in its vicinity (see Appendix).
This leads to an additional depolarization of
the $\mu$SR signal arising from the muons stopping in the pressure cell. Therefore,
the $\mu$SR asymmetry time spectra are characterized by
two components and may be described by the following expression:
\begin{align}\label{eq:AnFunc}
A(t) =& A_1 \cdot \exp\left(-\frac{1}{2}(\sigma^2+\sigma_n^2)t^2\right)\cos(\gamma_\mu B_1 t +\phi) +  \\ \nonumber
&A_2 \cdot \int P(B')\cos(\gamma_\mu B' t + \phi)dB'.
\end{align}
Here, $A_1$ and $A_2$ are  the initial asymmetries of the two components of the $\mu$SR
signal ($A_1$: sample, $A_2$: pressure cell), $\gamma_{\mu}$ is
the gyromagnetic ratio of the muon ($\gamma_\mu = 2\pi\times 135.5342$~MHz/T), and
 $\phi$ is the initial phase of the muon spin polarization. $B_1$ is the field in the center of the sample
(or approximately the mean field in the sample).
The parameter $\sigma$ denotes the muon
depolarization in the sample due to the field distribution created by the vortex lattice,
while $\sigma_n = 0.10(2)$~$\mu$s$^{-1}$ is a temperature, doping, and pressure
independent depolarization rate due to the nuclear moments present in the sample.
The total asymmetry is  $A_1 + A_2 = 0.275$ at 0.1~T and 0.265 at 0.5 T with $A_1/(A_1+A_2)\simeq 0.4$
($\simeq40$\% of the muon ensemble are stopping inside the sample).
$P(B')$ represents the magnetic field distribution probed by the muons stopping in the pressure cell as
described in detail in the Appendix.

\begin{figure}[!tb]
\includegraphics[width=0.95\linewidth]{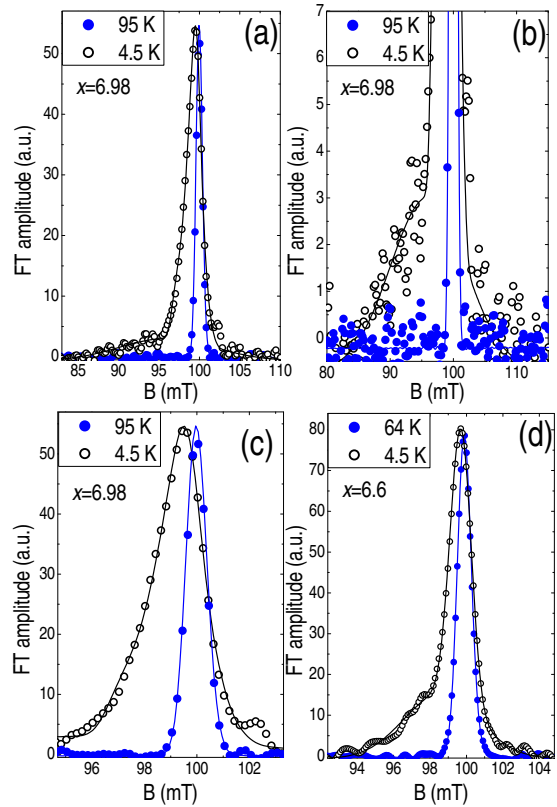}
\caption{(Color online) Fourier transform (FT) amplitude as a function of field for the spectra shown in
Fig.~\ref{fig:Fig1asy} [panels (a), (b), and (c)]. Panel (b) is the expanded [along $y-$axis]
view of panel (a) to show the signal from the sample.
Panel (c) is the expanded [along $x-$axis] view of panel (a)  to show the signal of the pressure cell.
Panel (d) shows the FT of the sample with $x=6.6$ below
and above $T_{\rm c} = 60$ K. The solid lines are the FTs of the fitted curves shown in
Fig. \ref{fig:Fig1asy}. The FT spectra  are slightly broadened due to a FT apodization
of 4~$\mu$s$^{-1}$. \label{fig:Fig2PB}}
\end{figure}

Figure \ref{fig:Fig1asy} exhibits $\mu$SR asymmetry time spectra of YBa$_2$Cu$_3$O$_{6.98}$
above ($T = 95$~K) and below ($T = 4.5$~K) the superconducting transition temperature
$T_{\rm c} = 89.6$~K obtained in an applied field of 0.1~T.
For a better visualization the spectra and the fits are shown in a rotating reference frame of 0.08~T.
Above $T_{\rm c}$ only a weak depolarization of the muon spin polarization is visible,\cite{Uemura88}
while below $T_{\rm c}$ the strong relaxation
of the $\mu$SR signal reflects the formation of the vortex lattice in the superconducting
state.\cite{Brandt88, Harschman87, Uemura88, Pumpin90, Sonier00RMP}
Figures~\ref{fig:Fig2PB}a, b, and c show the Fourier transforms (FT) of the $\mu$SR time spectra
shown in Fig. \ref{fig:Fig1asy}. In Fig. \ref{fig:Fig2PB}d
the FT spectra of YBa$_2$Cu$_3$O$_{6.6}$ below and above $T_{\rm c} = 60$ K are also shown.
The narrow signal around $B_{\rm app} = 0.1$~T in Fig. \ref{fig:Fig2PB}b originates from the pressure
cell, while the broad signal with a first moment significantly lower than $B_\mathrm{app}$ arises from
the superconducting sample. It can be seen that the signal of the pressure cell is also modified below
$T_{\rm c}$ due to the diamagnetic response of the superconducting sample.
The solid lines are the FTs of the fits to the data using Eq.~(\ref{eq:AnFunc}) (see also Appendix).
The good agreement between the fits and the data demonstrates that the model used here
describes the data rather well.

The whole temperature dependence of the $\mu$SR asymmetry time spectra was fitted globally with
the common parameters $B_{\rm app}$, $A_1$, $A_2$, and $\sigma_n$. Solely
the parameters $B_1$ and $\sigma$ were considered as temperature dependent free parameters.
As shown in the Appendix the field in the sample is macroscopically inhomogeneous due to the inhomogeneity of
demagnetization effects. $B_1$ is the field at the point $x=y=z=0$ ({\it i.e.,} the center of the sample).
In addition, the parameters describing the muon stopping distribution $x_{0,i}$ and $\sigma_i$
were kept the same for each temperature scan (see Eqs. (\ref{eq:stopDistr}) and (\ref{eq:PB3D}) in the Appendix).

\begin{figure}[tb]
\includegraphics[width=0.95\linewidth]{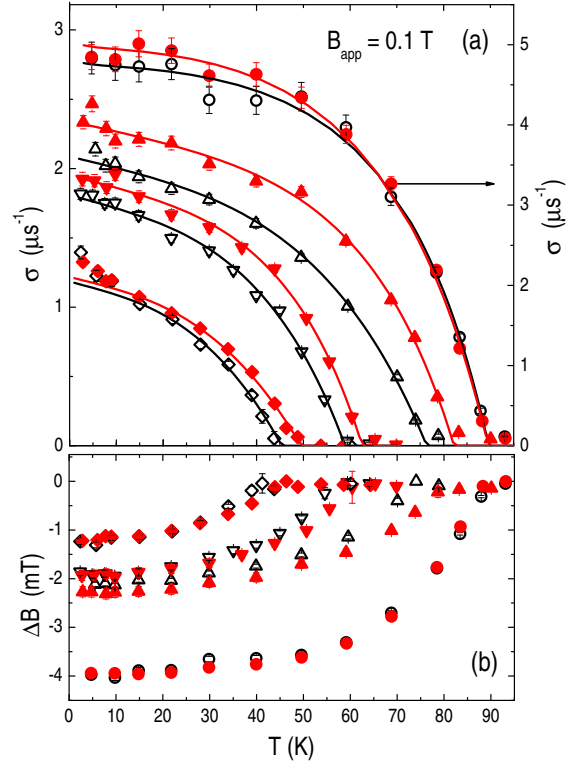}
\caption{(Color online) (a) Temperature dependence of $\sigma$ of YBa$_{2}$Cu$_{3}$O$_{x}$
measured at $B_\mathrm{app}=0.1$~T at
zero and applied hydrostatic pressures for  $x = 6.45$ ($\lozenge$: $P=0$ GPa; \txtr{$\blacklozenge$}: $P=1.1$ GPa),
$x = 6.6$ ($\triangledown$: $P=0$ GPa; \txtr{$\blacktriangledown$}: $P=1.1$ GPa), $x = 6.8$ ($\vartriangle$: $P=0$ GPa;
\txtr{$\blacktriangle$}: $P=1.1$ GPa),
and $x = 6.98$ ($\circ$: $P=0$ GPa;  \txtr{$\bullet$}: $P=1.4$ GPa).
The data were analyzed with Eq.~(\ref{eq:AnFunc}). The solid curves are fits to the
data with Eq.~(\ref{eq:SFLD}). (b) Diamagnetic shift of the field
$\Delta B = B_1-B_\mathrm{app}$ in the corresponding samples. $B_1$ is the mean field in the centrer
of the sample (see text and Appendix).
\label{fig:SFLD0p1}}
\end{figure}

\begin{figure}[tb]
\includegraphics[width=0.95\linewidth]{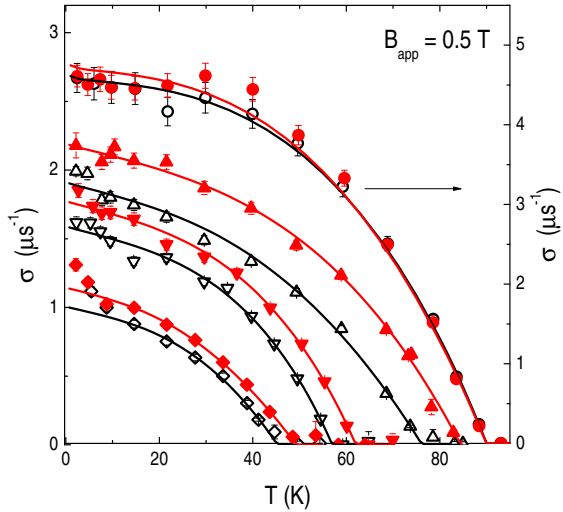}
\caption{(Color online) Temperature dependence of $\sigma$ of  YBa$_{2}$Cu$_{3}$O$_{x}$ measured at
$B_{\rm app} = 0.5$~T. The meaning of the symbols and the solid lines are the same as in
Fig. \ref{fig:SFLD0p1}(a).
\label{fig:SFLD0p5}}
\end{figure}

%
\begin{table}[!tb]
\caption[~]{Summary of the results obtained from the temperature dependence of $\sigma$ at 0.1 and 0.5~T
in YBa$_{2}$Cu$_{3}$O$_{x}$ using Eq. (\ref{eq:SFLD}). Note that for the sample with $x=6.45$ a
precise analysis of $\sigma$ was not possible due to the occurrence of spin-glass magnetism below $T\simeq$15~K.
Hence, the errors of these values of $\sigma(0)$ are rather large. }\label{table1}
\begin{center}
\begin{tabularx}{0.95\linewidth}{XXXXXXX }
\hline
\hline
$x$  &$P$    &$B_\mathrm{app}$   \hspace{0.3cm}   & $T_{\rm c}$  & $\sigma(0)$  & ${~\Delta_0}$  & ~$\Gamma_u$ \\
     &(GPa)     &(T)                   &  (K)   & ($\mu$s$^{-1}$)  &  $\overline{k_BT_{\rm c}}$   & (K)  \vspace{0.1cm}  \\
\hline
6.98 &0  &0.1      & 89.6(4)        &      4.76(7)      &    3.87(12)    &   ~15(5)          \\
6.98&1.4 &0.1      & 89.5(4)        &      4.97(7)    &     3.60(7)     &   ~15(5)          \\
6.98&0   &0.5     & 90.0(2)        &      4.56(7)      &    2.95(10)    &   ~15(5)          \\
6.98&1.4 &0.5    & 89.9(1)        &      4.72(7)    &    2.82(7)     &   ~15(5)      \vspace{0.1cm}    \\
6.8&0   &0.1   & 77.1(3)        &      2.07(5)        &   3.02(12)   &   ~0          \\
6.8&1.1 &0.1    & 83.2(5)        &      2.33(5)       &  3.48(15)     &   ~0          \\
6.8&0   &0.5   & 76.4(3)        &      1.91(5)        &   2.59(9)   &   ~0          \\
6.8&1.1  &0.5    & 82.3(5)        &     2.19(5)       &  2.80(8)     &   ~0       \vspace{0.1cm}   \\
6.6&0    &0.1   & 58.9(6)        &      1.79(5)        &   3.02(12)   &   ~0          \\
6.6&1.1  &0.1   & 62.6(5)        &      1.95(5)       &   3.27(12)    &   ~0          \\
6.6&0    &0.5   & 57.3(6)        &      1.58(5)        &   2.92(12)   &   ~0          \\
6.6&1.1  &0.5    & 62.3(6)        &      1.77(5)       &   2.89(11)    &   ~0        \vspace{0.1cm}  \\
6.45&0   &0.1     & 45.4(3)        &      1.17(7)        &    3.0(5)  &   ~0          \\
6.45&1.1 &0.1     & 49.5(5)        &      1.22(7)       &    3.0(5)   &   ~0          \\
6.45&0   &0.5    & 45.1(2)        &      1.00(7)        &     2.5(2)  &   ~0          \\
6.45&1.1  &0.5    & 48.7(2)        &      1.14(7)       &    2.5(2)   &   ~0          \\
\hline
\hline
\end{tabularx}
\end{center}
\end{table}
The temperature dependence of the depolarization rates $\sigma$
for $x = 6.98$, 6.8, 6.6, and 6.45 at $B_\mathrm{app}=0.1$
and 0.5~T obtained with Eq.~(\ref{eq:AnFunc}) are shown in Figs. \ref{fig:SFLD0p1} and \ref{fig:SFLD0p5},
respectively. The black empty points correspond
to the data measured at zero pressure, while the full red points correspond to the data measured
at 1.1~GPa (for $x=6.45$, 6.6, and 6.8) and 1.4~GPa (for $x=6.98$). The values of $T_{\rm c}$ and $\sigma(0)$
are in good agreement with previous results.\cite{Uemura88,Uemura89,Zimmermann95,Tallon95}
It is known that the order parameter in YBa$_2$Cu$_3$O$_{6.98}$ has
predominantly  the form of $\Delta = \Delta_0 (\hat{p_x^2}-\hat{p_y^2})$
[$\hat{p}_i=p_i/|\vec{p}|$ denotes component of the unit momentum vector in the reciprocal space along
the $i$-th axis].\cite{Khasanov07_multigap123,Khasanov07_multigap124,Lee06_RMP}
This implies a linear temperature dependence of the superfluid density $\rho_s$ down to very low temperatures due
to quasiparticle excitations at the gapless line nodes in the $\hat{p}_x = \pm |\hat{p}_y|$
directions on the Fermi surface.\cite{Sonier00RMP}
However, in Fig.~\ref{fig:SFLD0p1} we clearly see that $\sigma(T)$ tends to saturate
at low temperatures for
YBa$_2$Cu$_3$O$_{6.98}$ for both applied magnetic fields. Such a behavior was often observed in
$\mu$SR studies of polycrystalline samples\cite{Harschman87,Pumpin90} and was explained as
originating from a strong scattering of
electrons on impurities.\cite{Choi89,Hirschfeld93,Kim94,Xu95,Ohishi07}
This scattering can
strongly influence the temperature dependence of $\rho_s$, but it has a minor effect on the
superconducting transition temperature $T_{\rm c}$.  In previous theoretical works it was suggested
that such a behavior indicates scattering in the unitary limit.\cite{Hirschfeld93,Kim94}
Thus, the temperature
dependence of the superfluid density $\rho_s$ was analyzed with the``dirty $d$-wave model" of the
BCS theory in the unitary limit of carrier scattering as described in Ref.~\onlinecite{Xu95}:
\begin{align}\label{eq:SFLD}
\rho_s\propto\frac{1}{\lambda_{ab}^2}= \frac{4\pi e^2 N_f (v_f^{ab})^2}{c^2}
\int_0^{2\pi} \frac{d\phi}{2\pi} \sum_{n=0}^{\infty}
 \frac{|\Delta(\phi)|^2}{(\tilde{\epsilon}^2_n + |\Delta(\phi)|^2)^{3/2}}.
\end{align}
Here, $\lambda_{ab}$ is the in-plane magnetic penetration depth,
$\Delta(\phi) = \Delta_0\cos(2\phi)\cdot g(t)$ ($t = T/T_{\rm c}$) is the 2D-gap-function, and
$\tilde{\epsilon}_n = Z(\epsilon_n)\epsilon_n$ are
impurity renormalized Matsubara frequencies: $\epsilon_n = (2n+1)\pi T$. $\Delta_0$ is
the maximum of the gap function on the Fermi surface and $g(t)$ represents the temperature
dependence of the gap with $g(0)=1$. The parameters $N_f$ and $v_f$ are the
density of states at the Fermi level and the Fermi velocity, respectively. The constant $e$ and $c$
represent the electron charge and the speed of light.
The coefficients $ Z(\epsilon_n)$ are:\cite{Xu95}
\begin{equation}\label{eq:Zn}
Z(\epsilon_n) = 1 + \Gamma_u\frac{D_n(\epsilon_n) Z(\epsilon_n)}{\cot^2(\delta_0) +
[D_n(\epsilon_n) \epsilon_n Z(\epsilon_n)]^2},
\end{equation}
with
\begin{equation}
D_n(\epsilon_n) = \left< \frac{1}{\sqrt{ Z(\epsilon_n)^2 \epsilon^2_n + |\Delta(p_f)|^2}} \right>_{p_f},
\end{equation}
and $\delta_0 = \pi/2$ in the unitary limit. The angular brackets $\left< ...\right>_{p_f}$ denote averaging
over the Fermi surface.
In order to find $Z(\epsilon_n)$ and $g(t)$, Eq. (\ref{eq:Zn}) is solved together with the following
equation:\cite{Xu95}
\begin{align}\label{eq:gt}
&\frac{1}{2\pi T}\left\{ \ln\left(\frac{T}{T_{\rm c}}\right)+\psi\left(\frac{1}{2}+\frac{\Gamma_u}{2\pi T}\right)
-\psi\left(\frac{1}{2}+\frac{\Gamma_u}{2\pi T_{\rm c}}\right) \right\} = \\ \nonumber
& \sum_{n=0}^\infty\left[\left< \frac{|e({p}_f)|^2}
{(Z(\epsilon_n)^2{\epsilon}^2_n + |\Delta(p_f)|^2)^{3/2}} \right>_{{p}_f}
-\frac{1}{\epsilon_n +\Gamma_u}\right].
\end{align}
Here, $\psi(x)$ is the digamma function. Note that the impurity scattering influences mainly
$\epsilon_n$ while the temperature dependence of the gap $g(t)$ changes only slightly for a reasonable
scattering rate $\Gamma_u$. In the clean limit ({\it i.e.}, $\Gamma_u = 0$ and $Z(\epsilon_n)=1$, $\forall \, n$)
the normalized function $g(t)$ is very close to the analytical approximations derived from BCS theory.\cite{tinkham}

Fits of Eq.~(\ref{eq:SFLD})  to $\sigma(T) \propto 1/\lambda_{\rm ab}(T)^2$
measured at various hydrostatic pressures are presented
in Figs.~\ref{fig:SFLD0p1} and \ref{fig:SFLD0p5}.
The corresponding values for $\Delta_0$, $T_{\rm c}$, $\sigma_0$, and $\Gamma_{u}$
obtained from the analysis
are summarized in Table~\ref{table1}.
The data for zero and applied pressure and the same doping $x$ were analyzed simultaneously with the
common parameter $\Gamma_u$ which characterizes the relaxation rate of the Cooper pairs on impurities.
As shown in Table~\ref{table1} the data for the underdoped samples ($x = 6.45$, 6.6, and 6.8)
are well described by the clean limit
$d$-wave model, while for the overdoped sample ($x=6.98$)  $\Gamma_u = 15(5)$ K.
Here, we note that all the studied samples originate from the same batch and
have an identical thermal history, except of the last process of the oxygen reduction.
Therefore, we cannot explain why only the sample with $x=6.98$ exhibits a saturation of $\sigma$ in
the low temperature limit and why it has such a high scattering rate $\Gamma_u = 15(5)$ K.
Consequently, we cannot exclude the possibility of a modification of the order parameter in overdoped
\YBCOx where the pseudogap state gradually vanishes.
Such a behavior was also observed previously in optimally
doped or overdoped polycrystalline  samples of \YBCOx.\cite{Zimmermann95,Pumpin90,Harschman87,Uemura88}
However, in single crystal \YBCOx close to optimum doping a linear temperature
dependence of $1/\lambda^2$ at low temperatures was also reported.\cite{Riseman95,Sonier00RMP}
For the sample with $x=6.45$ only the data above 15~K were analyzed, since below 15 K
the occurrence of field induced spin-glass magnetic order hinders a precise determination of $\sigma$.

\begin{figure}[tb]
\includegraphics[width=0.95\linewidth]{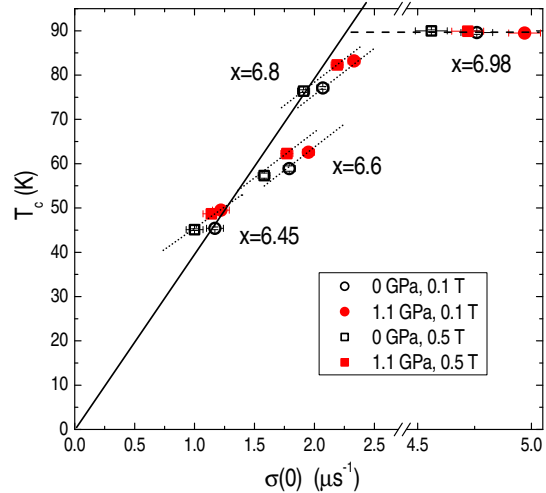}
\caption{(Color online) $T_{\rm c}$ vs. $\sigma(0)$ (Uemura plot) at zero and applied pressure
for YBa$_2$Cu$_3$O$_x$ with $x = 6.45$, 6.6, 6.8, and 6.98.
The solid line is the Uemura line while the dashed line is a guide to the eye.
The dotted lines represent the pressure effect on  $T_{\rm c}$ and $\sigma(0)$.
\label{fig:UemuraGap}}
\end{figure}

\begin{figure}[tb]
\includegraphics[width=0.95\linewidth]{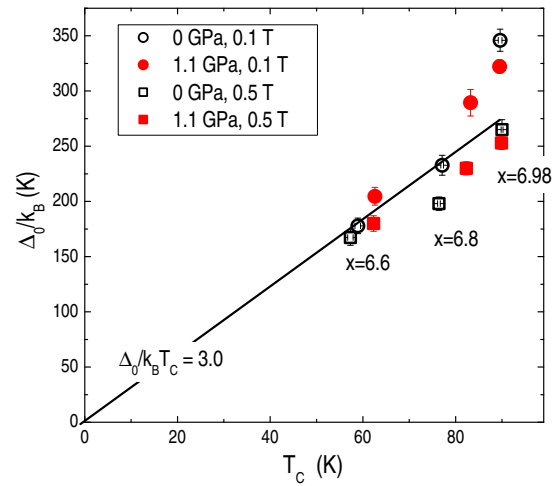}
\caption{(Color online)  Relation between $\Delta_0$ and $T_{\rm c}$ for
YBa$_2$Cu$_3$O$_x$ with $x = 6.6$, 6.8, and 6.98. The solid line
corresponds to $\Delta_0/k_B T_{\rm c} = 3$ (weak-coupling BSC superconductor: $\Delta_0/k_B T_{\rm c} = 1.76$).
Both $\Delta_0$ and $\Delta_0/k_B T_{\rm c}$ increase with increasing pressure.
\label{fig:DeltaVsSg}}
\end{figure}

\section{Discussion}

\begin{figure}[tb]
\includegraphics[width=0.95\linewidth]{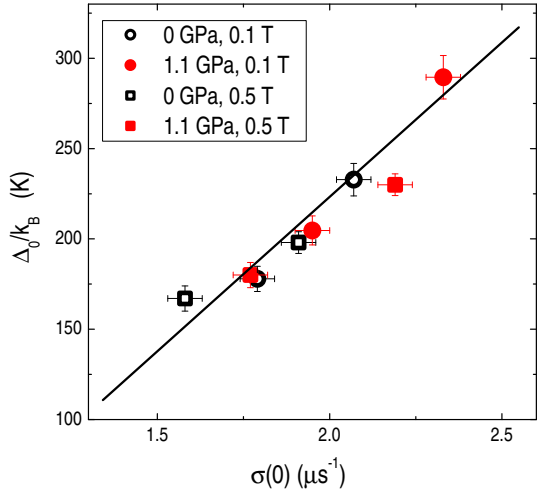}
\caption{(Color online)
The gap $\Delta_0$ as a function of $\sigma(0)$ for the underdoped samples
of YBa$_2$Cu$_3$O$_x$ with $x=6.6$ and 6.8.
The linear relation between $\sigma(0)$ and $\Delta_0$ is better
fulfilled under hydrostatic pressure than the Uemura relation
$T_{\rm c}$ vs $\sigma(0)$ and  $\Delta_0/k_B T_{\rm c}$ vs. $T_{\rm c}$
(see Figs. \ref{fig:UemuraGap} and \ref{fig:DeltaVsSg}).
The line is a guide to the eye.
\label{fig:DeltaVsSg2}}
\end{figure}

\begin{table}[!tb]
\caption[~]{Values of $\alpha_{\rm p} = (\partial T_{\rm c}/\partial P)/(\partial \sigma/\partial P)$
for the underdoped \YBCOx samples investigated in this work ($x=6.45$, 6.6, and 6.8). }\label{table2}
\begin{center}
\begin{tabularx}{0.95\linewidth}{XXXXXXX }
\hline
\hline
$x$ & $\alpha_{\rm p}$ (K/$\mu$s$^{-1}$)  &$\alpha_{\rm p}$ (K/$\mu$s$^{-1}$)   \\
     &0.1 T     &0.5 T                \\
\hline
6.45 &--  & 25(18)          \\
6.6 &23(11)  & 26(6)          \\
6.8 &23(7)  & 21(6)          \\
\hline
\hline
\end{tabularx}
\end{center}
\end{table}

The main subject of the present study is the pressure effect on the superconducting gap $\Delta_0$
and the superfluid density $\rho_s \propto \sigma$. The Uemura relation\cite{Uemura89}, implying
the linear relation between $T_{\rm c}$ and $\rho_s$ for underdoped
cuprate superconductors, was established soon after the discovery of HTS\cite{Bednorz86} and is one of the important
criteria which a microscopic theory of HTS should explain. 
The Uemura relation for the data summarized in Table \ref{table1} is shown in Fig. \ref{fig:UemuraGap}.
As indicated by the dotted lines the slope
$\alpha_{\rm p} = (\partial T_{\rm c}/\partial P)/(\partial \sigma/\partial P)$ is systematically
smaller than that suggested by the Uemura line with $\alpha_U=\partial T_{\rm c}/\partial \sigma \simeq 40$
K/$\mu$s$^{-1}$.
The values of $\alpha_{\rm p}$ for the underdoped samples investiganted in this work are summarized in
Table \ref{table2}.  Note that due to magnetism below $\sim 15$~K the error of $\sigma(0)$ for
the sample with $x=6.45$ is rather large.
The weighted mean value of $\alpha_{\rm p} \simeq 23(4)$~K/$\mu$s$^{-1}$
is a factor of $\simeq 2$ smaller than $\alpha_U \simeq 40$ (K$\mu$s$^{-1}$).
Such a substantial deviation from the Uemura line (with a lower value of $\alpha_{\rm p}$) was also observed
by pressure experiments in YBa$_2$Cu$_4$O$_8$ using a magnetization technique.\cite{Khasanov05_P124}
This is in contrast to pressure effect results obtained for the organic superconductor $\kappa$-(BEDT-TTF)$_2$Cu(NCS)$_2$
which follow the Uemura relation.\cite{Larkin01}
Interestingly, a slope with a factor two smaller than that of the Uemura line was also found by OIE
studies of cuprate superconductors.\cite{Khasanov03_OIE_JPCM}
This suggests a strong influence of pressure on the lattice dynamics.
%
It is known that the pressure dependence of the superconducting transition temperature is determined by two
mechanisms: (i) The pressure induced charge transfer to CuO$_2$ planes $\Delta n_h$
and (ii) the pairing interaction  $V_\mathrm{eff}$ which depends on
pressure.\cite{Chen00,Sahiner99,Gupta95,Almasan92,Neumeier93,Angilella96,Mello97,Sarkar98}

For the underdoped samples the former mechanism dominates (85-90\%)
the pressure effect on $T_{\rm c}$.\cite{Gupta95,Almasan92,Chen00}
Therefore, one can separate the pressure effect on  $\sigma$
also in two components $\Delta \sigma = \Delta\sigma_\mathrm{ch}+\Delta\sigma_V$.
The first term $\Delta \sigma_\mathrm{ch} \simeq (1/\alpha_U) (\partial T_{\rm c}/\partial P) P$
follows the Uemura line and is mainly due to the charge transfer to the plane.
The second term $\Delta \sigma_V \simeq (1/\alpha_{\rm p}-1/\alpha_U)(\partial T_{\rm c}/\partial P) P$
describes the increase of the superfluid density solely due to a change of the pairing interaction.
This increase of the superfluid density is equivalent to a decrease of the effective mass of the superconducting carriers,
since $\Delta \sigma_V/\sigma=\Delta  \lambda_V^{-2}/\lambda^{-2} = -\Delta m^*_V/m^*$.\cite{Khasanov05_P124}
Therefore, the pressure-induced change of the effective carrier mass can be written as:
\begin{align}\label{eq:PressEff}
d \ln(m^*_V)/dP&=-d \ln(\lambda^{-2}_V)/dP\equiv-(\Delta \sigma_V /\sigma)/\Delta P\\ \nonumber
& \simeq (\alpha_U/\alpha_{\rm p}-1)(\partial T_{\rm c}/\partial P)/T_{\rm c} \\ \nonumber
&\simeq 3/T_{\rm c} \mathrm{ ~GPa^{-1}}.
\end{align}
Here, $T_{\rm c}$ and $\sigma$ are taken at zero pressure and the value of
$(\partial T_{\rm c}/\partial P)\simeq 4$~K/GPa
was used. This value is practically doping independent in underdoped
\YBCOx  for $6.45\le x\le 6.8$.\cite{Gupta95}
The quantity $\Delta \lambda^{-2}_V$ describes the change of the superfluid density solely
due to a modification of the pairing interaction $V_\mathrm{eff}$ by pressure.
It is remarkable to observe the qualitative agreement between $d \ln(\lambda^{-2}_V)/dP$ and that found in OIE
studies for $d \ln\lambda/d \ln M_O$ at different carrier dopings
($d \ln M_O$ is the relative change of oxygen mass).\cite{Khasanov03_OIE_JPCM}
Indeed, Eq. (\ref{eq:PressEff}) predicts that the pressure effect on $m*_V$ strongly increases
with decreasing $T_{\rm c}$.

Another interesting result is the quite small pressure dependence of $\sigma$ in the overdoped
sample with $x=6.98$, which is approximately a factor of $\simeq 2$ weaker than that reported
from magnetization measurements.\cite{DiCastro09}
In Fig. \ref{fig:DeltaVsSg} the gap magnitudes $\Delta_0$ for the samples with $x=6.6$, 6.8, and 6.98
are plotted as a function of $T_{\rm c}$.
For the underdoped samples ($x = 6.6$ and 6.8) both $\Delta_0$ and $\Delta_0/k_BT_{\rm c}$ increase
upon increasing applied pressure.
This suggests an increase of the coupling strength with increasing pressure.
This behavior is different from that found for the OIE on $\Delta_0$,
 where a proportionality between $\Delta_0$ and $T_{\rm c}$ was
found, implying a constant ratio of $\Delta_0/k_BT_{\rm c}$.\cite{Khasanov08_OIE_gap}
In the overdoped sample ($x=6.98$), Eq. (\ref{eq:SFLD})
suggests a small reduction of the coupling strength with
increasing pressure. However, as was mentioned above, the absence of a linear temperature dependence
of $\sigma$ at low temperatures for the sample with $x=6.98$ might also indicate that
the superconducting order parameter is not of purely $d$-wave
character.\cite{Khasanov07_multigap124,Khasanov07_multigap123} This, on the other hand,
may influence the result for $\Delta_0$ and its pressure dependence.

%

In Fig. \ref{fig:DeltaVsSg2} for the underdoped samples ($x=6.6$ and 6.8) $\Delta_0$ is plotted
vs. $\sigma(0)$, showing a linear correlation between the two quantities. Note, that this correlation
does not change with the application of hydrostatic pressure.
This is in contrast to what is observed for the Uemura relation $T_{\rm c}$ vs. $\sigma(0)$ and
$\Delta_0/k_B T_{\rm c}$ vs. $T_{\rm c}$
(see Figs. \ref{fig:UemuraGap} and \ref{fig:DeltaVsSg}).

\section{Conclusions}

The pressure dependence of the magnetic penetration depth $\lambda$ of
polycrystalline YBa$_{2}$Cu$_{3}$O$_x$ ($x = 6.45$, 6.6, 6.8, and 6.98)
was studied by $\mu$SR.  The pressure dependence of the superfluid density
$\rho_s \propto \sigma \propto 1/\lambda^2$ as a function of the superconducting transition $T_{\rm c}$
temperature does not follow the well-known Uemura relation.\cite{Uemura89} The ratio
$\alpha_{\rm p} = (\partial T_{\rm c}/\partial P)/(\partial \sigma/\partial P) \simeq 23(4)$ K/$\mu$s$^{-1}$
is a factor of $\simeq 2$ smaller than that of the Uemura relation observed for underdoped samples.
However, the value of $\alpha_{\rm p}$ is quite close  to that found
in OIE studies,\cite{Khasanov03_OIE_JPCM} indicating a strong influence of pressure on the lattice
degrees of freedom. We conclude that the contribution of carrier doping to the pressure dependence of $\lambda$
is similar to the OIE on $\lambda$.
A weak pressure dependence of the superfluid density $\rho_s$ was found in the overdoped sample ($x=6.98$).
The superconducting gap $\Delta_0$ and the BCS  ratio $\Delta_0/k_BT_{\rm c}$ both increase with increasing
applied hydrostatic pressure in the underdoped samples, implying an increase of the coupling strength with pressure.
Although the Uemura relation does not hold and the BCS ratio is increasing with pressure in underdoped
samples, the relation between $\Delta_0$ and the $\mu$SR relaxation rate $\sigma$ is invariant under pressure.
Finally, a model to analyze TF $\mu$SR spectra of magnetic/diamagnetic samples loaded into a pressure
cell was developed and successfully used in this paper (see Appendix),
resulting in a substantial reduction of the systematic errors in the data analysis.

\section*{Acknowledgements}
We are grateful to M.~Elender for his technical support during the experiment and D. Andreica
for providing the pressure cells.
This work was performed at the Swiss Muon Source (S$\mu$S), Paul
Scherrer Institut (PSI, Switzerland). We acknowledge support
by the Swiss National Science Foundation, the NCCR
{\it Materials with Novel Electronic Properties} (MaNEP), the SCOPES grant No. IZ73Z0-128242,
and the Georgian National Science Foundation grant GNSF/ST08/4-416.

\appendix*

\section{Field distribution in a pressure cell loaded with a sample
with a non-zero magnetization}

Samples with a strong magnetization placed in a pressure cell with an
applied magnetic field induce a magnetic field in the space around the sample.
Typical examples of such samples are superconductors (strong diamagnets), superparamagnets,
and ferro- or ferrimagnets. Thus,
muons stopping in a pressure cell (PC) containing the sample will undergo precession in the vector sum of
the applied field and the field induced by the sample. This spatially inhomogeneous field leads to
an additional depolarization of the muon spin polarization which
depends on the applied field and the induced field together
with the spatial stopping distribution of the muons.

\begin{figure}[tb]
\includegraphics[width=0.52\linewidth]{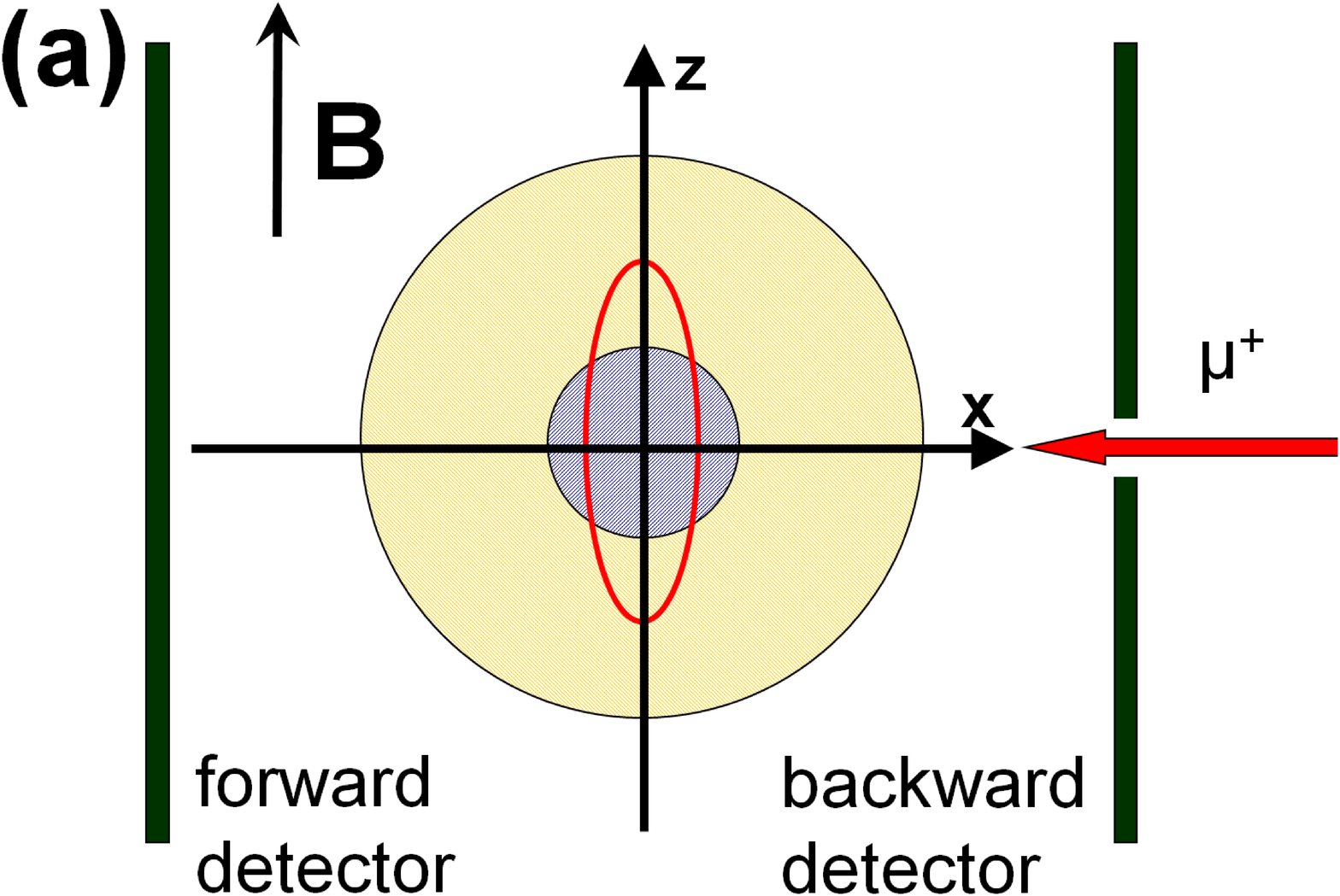}
\includegraphics[width=0.43\linewidth]{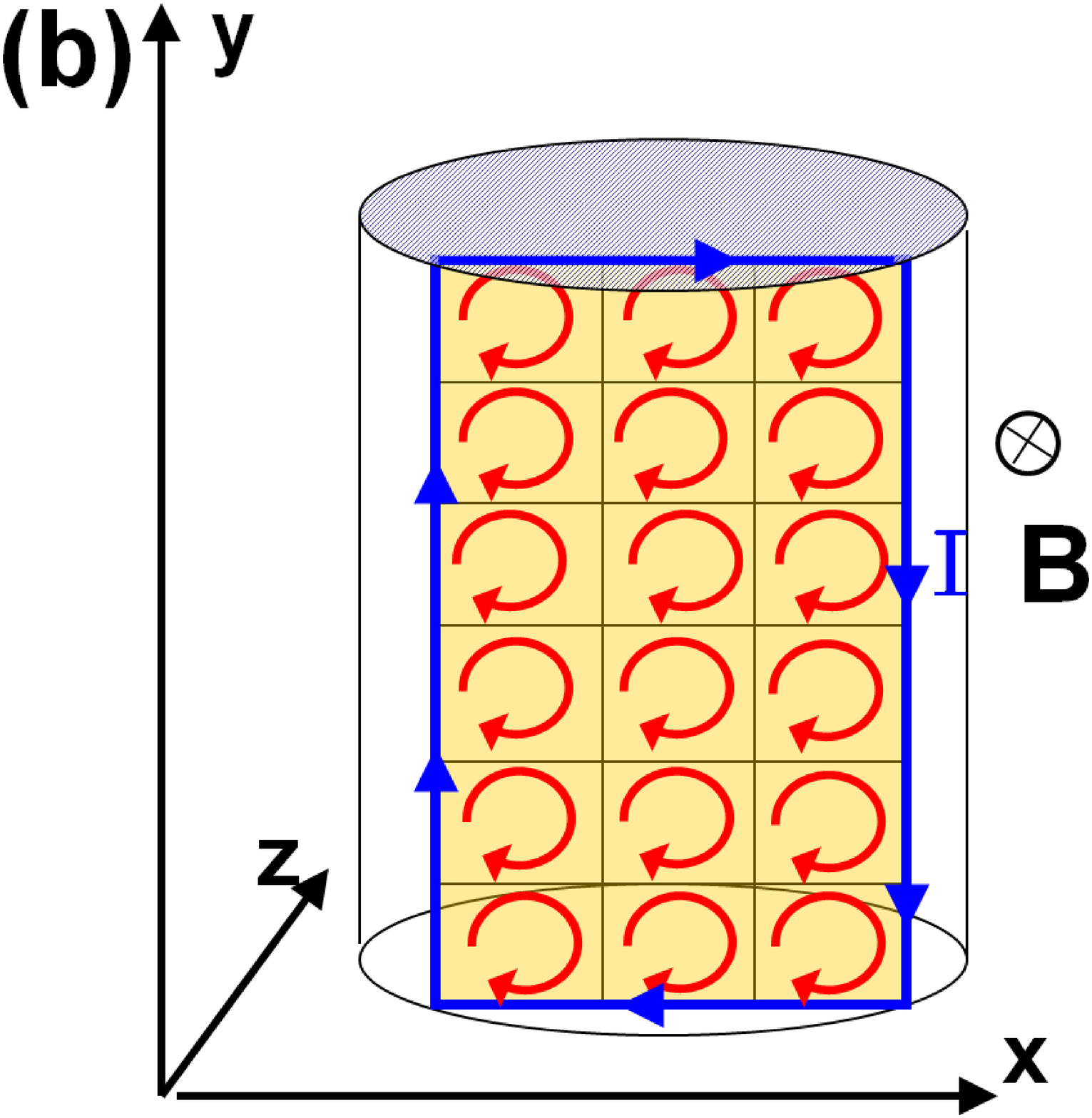}
\includegraphics[width=0.49\linewidth]{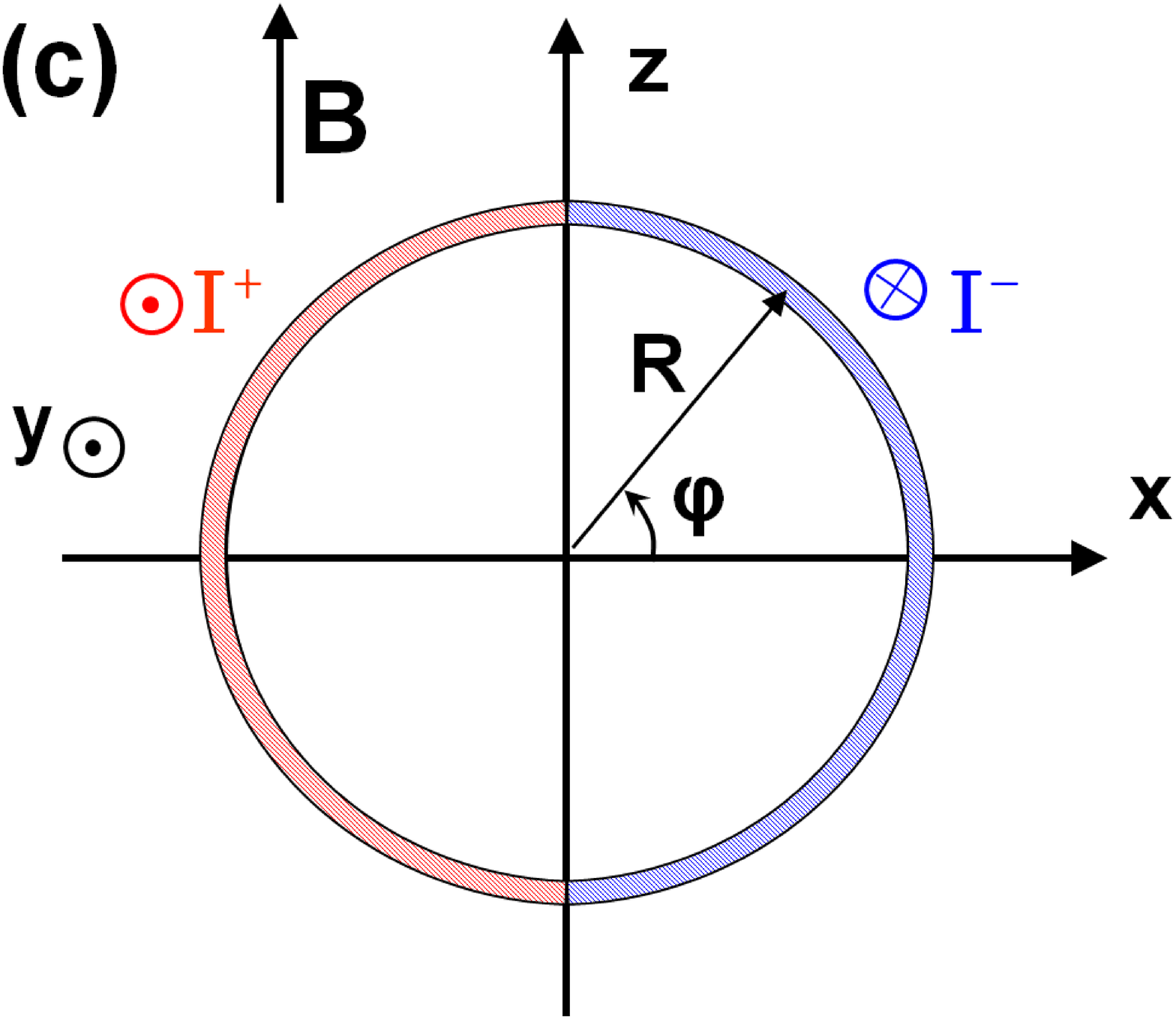}
\includegraphics[width=0.43\linewidth]{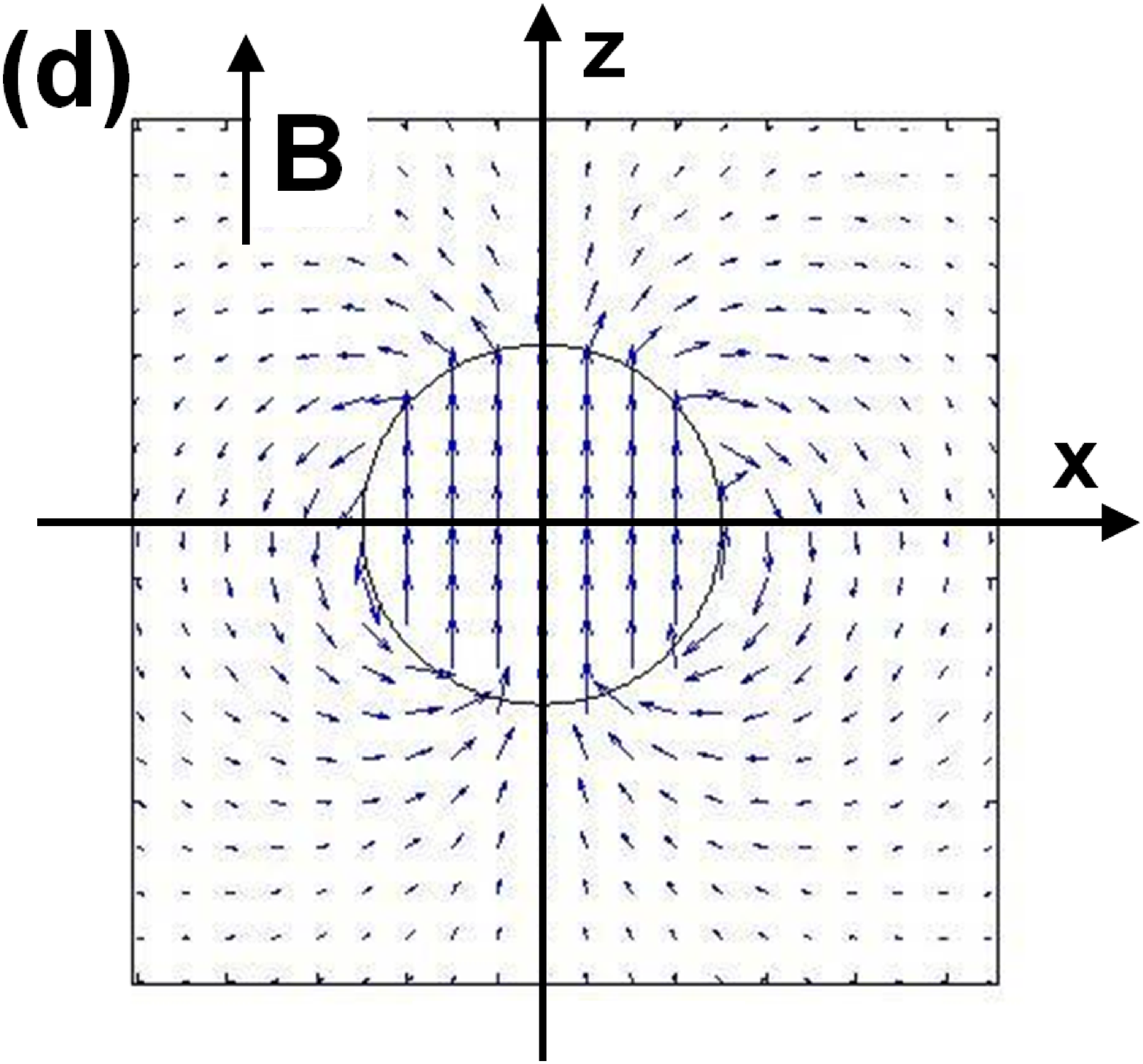}
\caption{(Color online)
(a) Schematic sketch of the $\mu$SR pressure instrument GPD at the Paul Scherrer
Institute: Cylindrical sample (blue); pressure cell (yellow); muon stopping distribution
(red ellipse), and forward and backward positron detectors (black).
(b) Illustration of the surface current on a slice of a homogeneously
magnetized cylindrical sample. The magnetic field induced by this slice is equivalent
to the magnetic field of the surface currents. (c) Cross section of the cylindrical
sample and the surface current distribution in $xz-$plane. (d) Magnetic
field map of the surface currents as illustrated in panels (b) and (c).
\label{fig:SampleInPcell}}
\end{figure}

Consider the most simplest case of a sample with the shape of a round cylinder of
hight $H$ and radius $R$ placed into a cylindrical pressure cell
with the same internal radius $R$ (Fig.~\ref{fig:SampleInPcell}a).
Typical pressure cell radii used for $\mu$SR studies are $R= 2.5$ - 4 mm.
In standard transverse field (TF) $\mu$SR experiments the pressure cell is
placed with the cylinder axis oriented vertically while the magnetic field is applied
perpendicular to the cylinder axis of the pressure cell and the muon beam direction (see Fig.~\ref{fig:SampleInPcell}).
Let us introduce a cartesian coordinate system with the $y$-axis along the sample cylinder axis,
and the $z$-axis
along the direction of the applied field. Thus, the $x$-axis is along the initial
muon beam direction which is perpendicular to the forward and backward detector planes (see Fig.~\ref{fig:SampleInPcell}).
The origin of the coordinate system is located in the center of the sample.

In an applied magnetic field ${\bf H}$ (along the $z$-direction) the sample has a magnetization ${\bf M}$.
This magnetization is the source of an induced field $\mathbf{H'}(\mathbf{r})$. Let us assume that
$\mathbf{H'}$ is much weaker that the applied field ${\bf H}$ which is the case
for superconductors in a magnetic field of $\mu_0 H\gg B_{c1}$ ($B_{c1}$ is the first critical field).
Thus, one can neglect the spatial variation of the magnetization due to the additional induced field:
$M = M(H + H'(\mathbf{r}))\simeq M(H)$.
Typically half (or even more) of all the muons are stopping in the
PC outside of the sample volume. The muons stopping in the macroscopically
inhomogeneous field of the PC contribute to an additional relaxation of the $\mu$SR signal.
In order to describe the total $\mu$SR time spectrum (sample and PC)
one has to model the field distribution ${\bf H'(r)}$. For an applied field $ H \gg {H'}(\mathbf{r})$
one can neglect the influence of $H'_x(\mathbf{r})$ and $H'_y(\mathbf{r})$ on the
$\mu$SR time spectrum, since only the $z$-component $H'_z(\mathbf{r})$ contributes
significantly to the muon depolarization.
The induced magnetic field  ${\bf H'(r)}$ created by a cylindrical sample can be calculated as follows:\cite{Feynman64}
\begin{equation}\label{eq:Binduced}
\mathbf{H'}(\mathbf{r}) = \frac{1}{4\pi}\int\limits_V \left[
\frac{3(\mathbf{M}\cdot(\mathbf{r}-\mathbf{r'}))\mathbf{(\mathbf{r}-\mathbf{r'})}}
{|\mathbf{r}-\mathbf{r'}|^5} -\frac{\mathbf{M}}{|\mathbf{r}-\mathbf{r'}|^3} \right] d\mathbf{r'}
\end{equation}
\begin{figure}[!tb]
\includegraphics[width=0.49\linewidth]{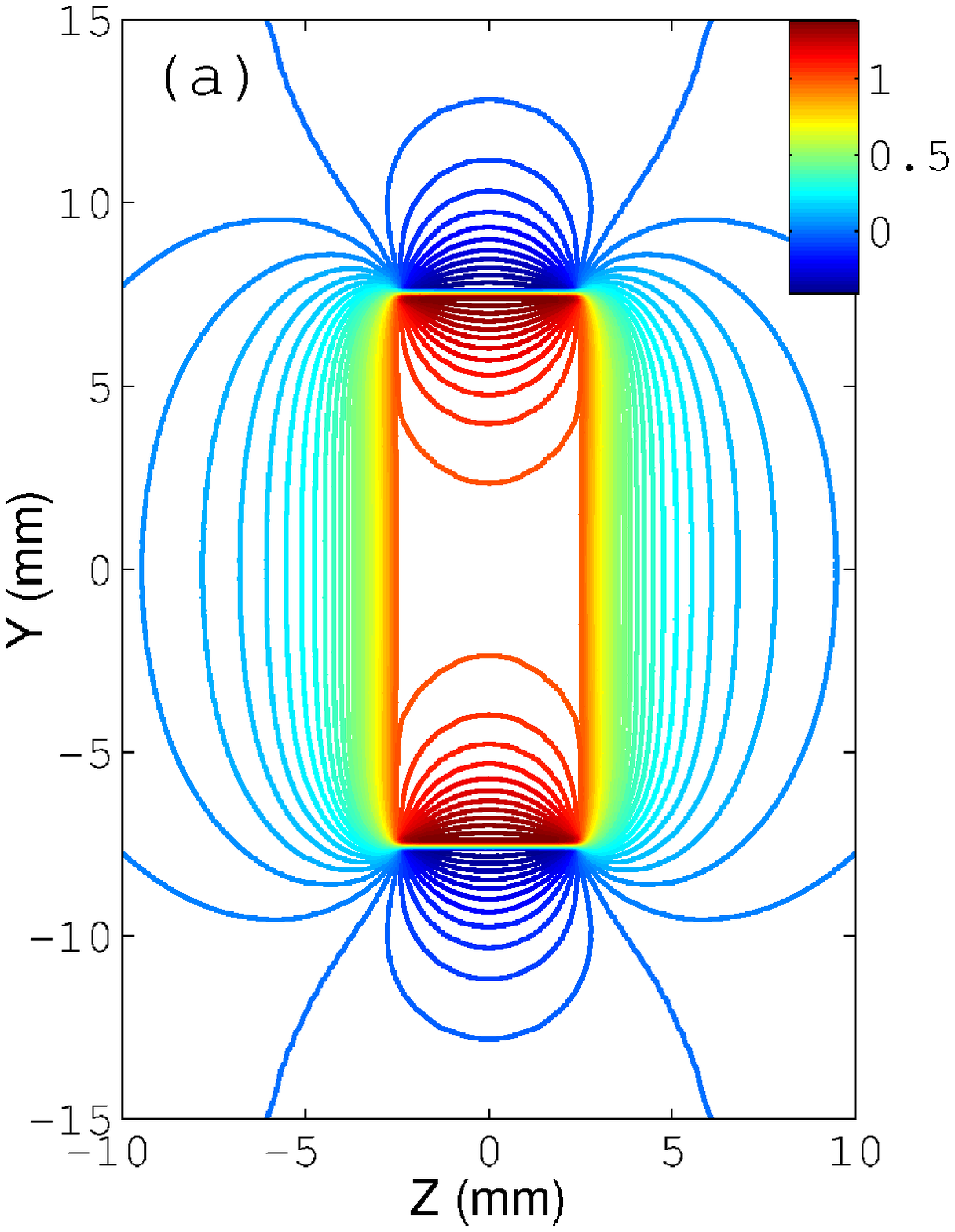} 
\includegraphics[width=0.47\linewidth]{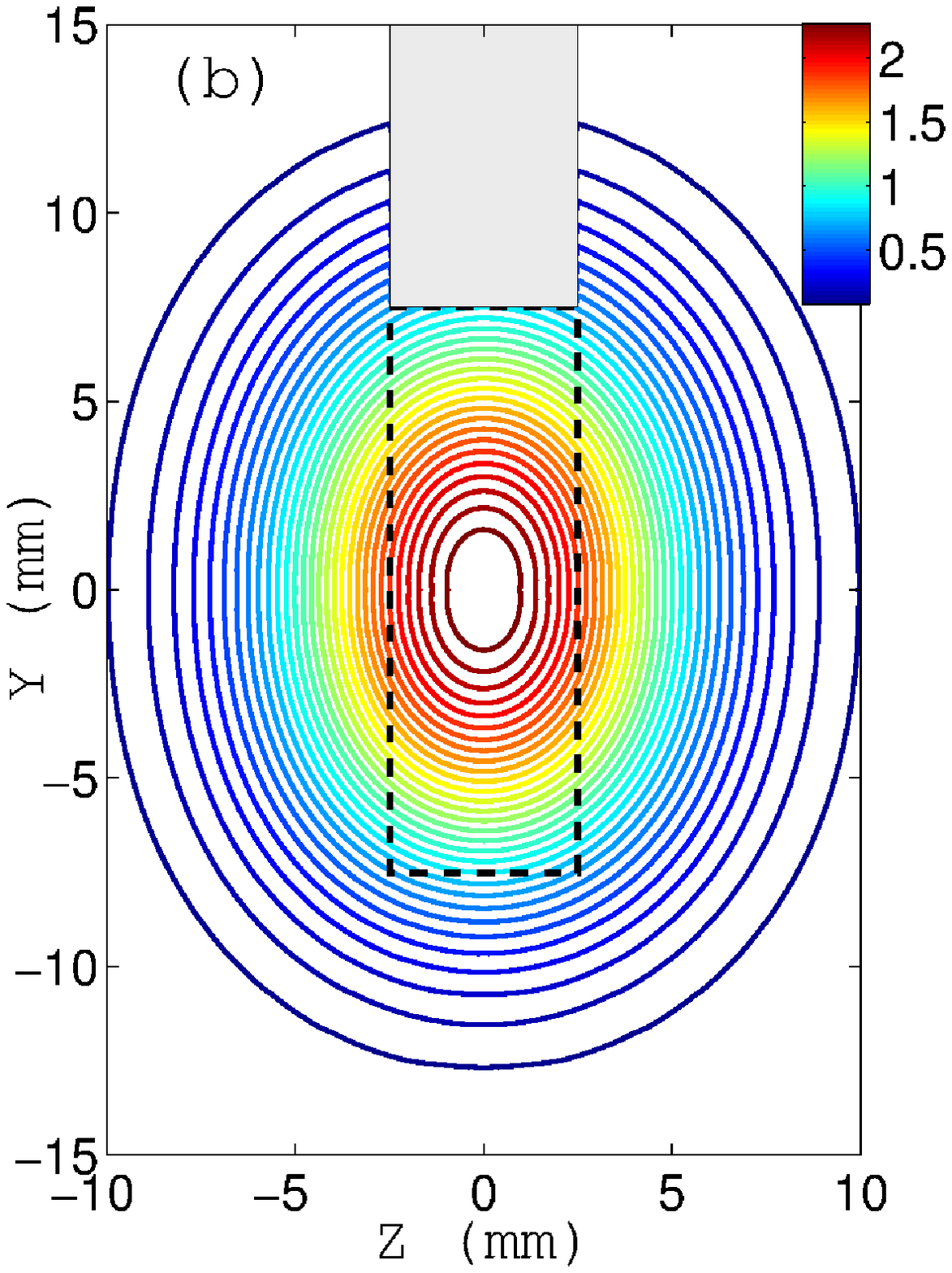}\\
\includegraphics[width=0.95\linewidth]{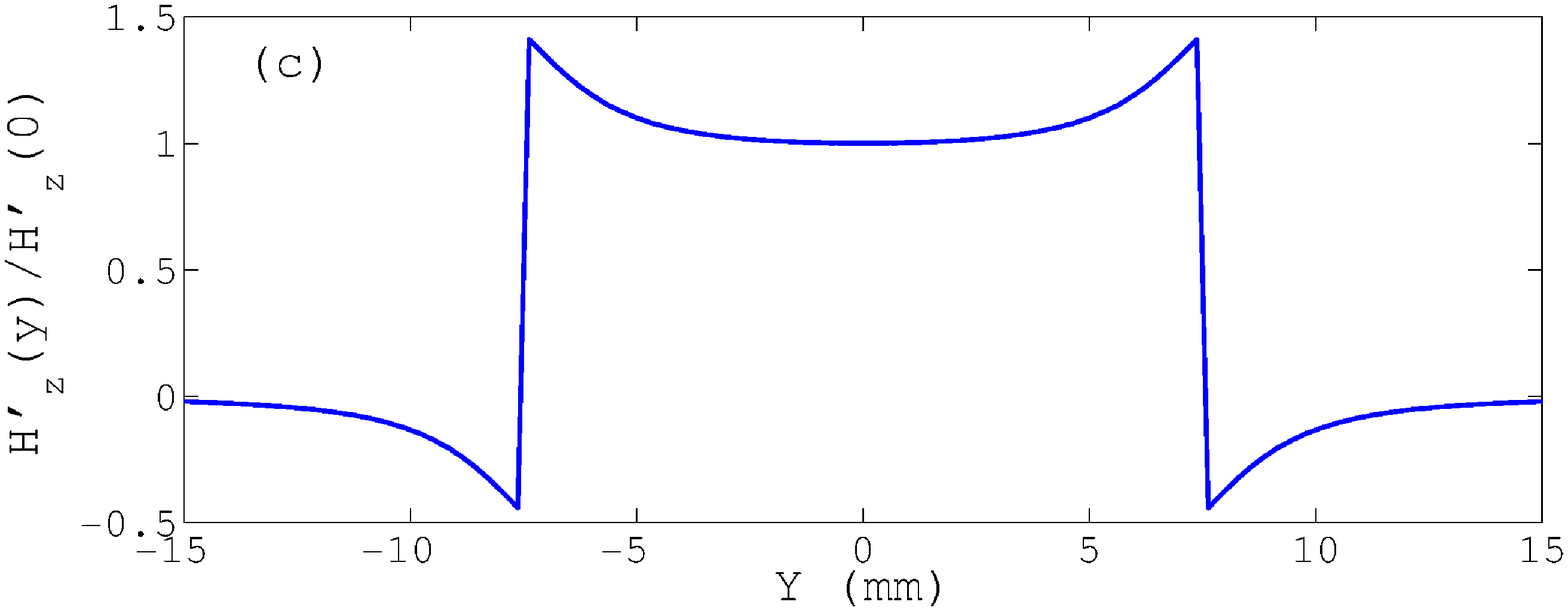}\\
\includegraphics[width=0.95\linewidth]{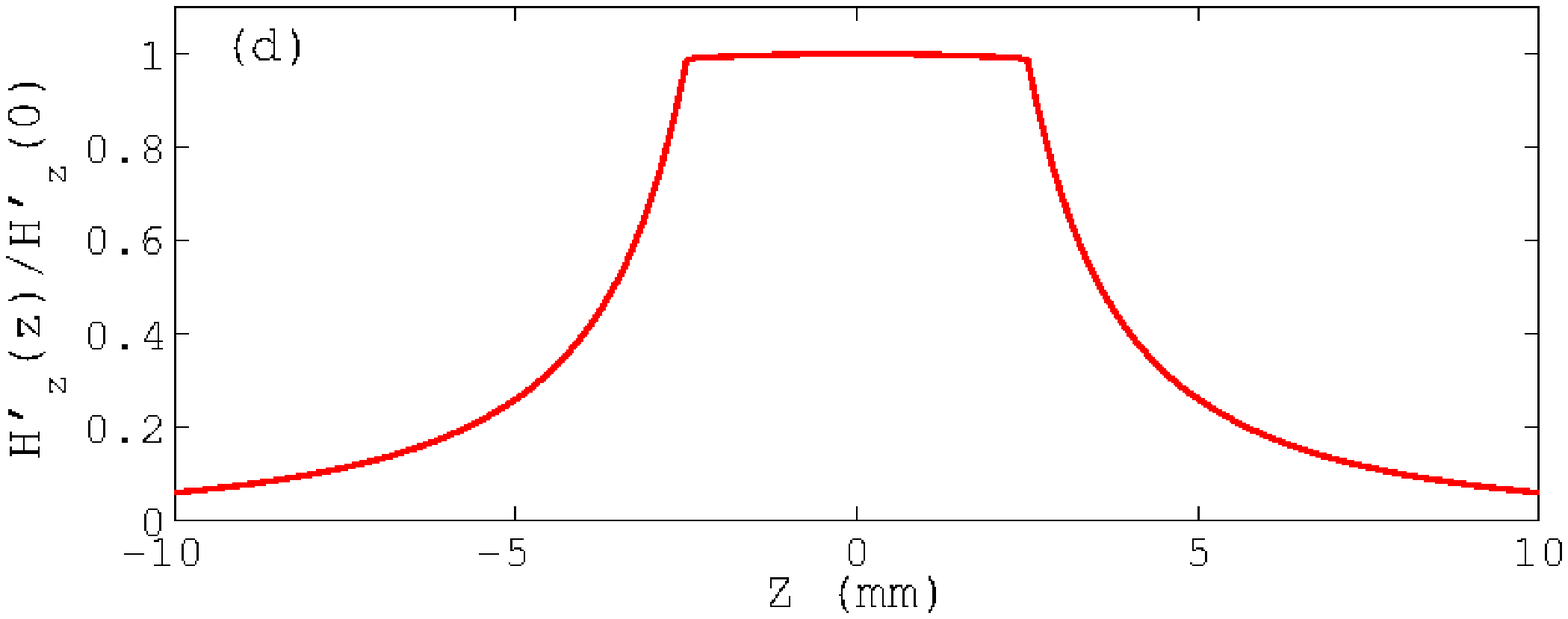}\\
\caption{(Color online) (a) Contour plot of the field distribution ${H'_z}(y,z)$ in the
$yz-$plane for a cylindrical sample with $R=2.5$ mm and $H=15$ mm (the geometry of the sample used in the experiment).
(b) Contour plot of the muon stopping
distribution in the $yz-$plane. The gray area on the top of the sample corresponds to the
empty pressure cell space  where no muons stop (this space is filled with a low-density
pressure transmission medium). The dashed line indicates the sample space.
(c) Magnetic field profile of ${H'_z}$ along the $y-$axis and (d) magnetic field profile
of ${H'_z}$ along the $z-$axis.
\label{fig:SampleInPcell2}}
\end{figure}
Here, the integral is taken over the sample volume V. For a sample with a constant magnetization
the three-dimensional integral can be replaced by surface integrals.
Let us take one slice of width $dz$ out of the sample cylinder and divide it into many
small squares $dA = dxdy$ (see Fig.~\ref{fig:SampleInPcell}b).
The field created by the elementary cell of volume $dV = dxdydz$ with magnetization
$M$ is equivalent to the field created by the current $I_z = M dz$ circulating within this
square slice as shown in Fig.~\ref{fig:SampleInPcell}a. It is obvious
that integration of this field over the whole slice volume will leave only a current $I_z$
flowing over the perimeter of the slice. The total field of the cylinder is the  integral of the
fields created by these slices with constant current $I_z$ (see Figs.~\ref{fig:SampleInPcell}b and c).

According to the law of Bio-Savart the field in a point ${\bf r}$ created
by the elementary currents I$d\mathbf{\ell}$ at the surface of the cylinder
(with coordinates ${\bf r_s}$) is:\cite{Feynman64}
\begin{equation}\label{eq:BioSavart}
{\bf H'}({\bf r}) = \oint\limits_{S}  \frac{ I}{4\pi} \frac{[d{\bf \ell_s\times (r-r_s)}]}{|\mathbf{r}-\mathbf{r_s}|^3}.
\end{equation}
The integration is taken over the surface S of the sample and $d{\bf \ell_s}$ is the elementary length on the
surface with its direction along the current (the subscript $s$ denotes quantities related to the surfaces
of the sample.

The spacial magnetic field distribution around the ferro/paramagnetic sample calculated with
Eq. (\ref{eq:BioSavart}) in $x$-$z$ plane is shown in Fig.~\ref{fig:SampleInPcell}d. The total field in
the pressure cell is the vector sum of this field and the homogeneous external field.
It is obvious from the figure that the field along the $z$-axis is
higher(lower) than the external field in a ferromagnet(diamagnet).
Along the $x$-axis, on the other hand, the field is
lower(higher) than the external field in a ferromagnet(diamagnet).
The maximal (minimal) induced field in the
PC are just on the border of the sample/pressure cell along $z$ ($x$) direction.
Note that demagnetization effects are naturally accounted for by using Eq. (\ref{eq:BioSavart}).
Since the sample is not elliptical this leads to field inhomogenieties within the volume of
the sample (see Fig. \ref{fig:SampleInPcell2}).
As an example Fig.~\ref{fig:SampleInPcell2} shows the magnetic field distribution in the $yz-$plane for a cylindrical
sample with $H=15$~mm and radius $R=2.5$ mm, together with fields along $z-$ and $y-$axes
calculated with Eq.~(\ref{eq:BioSavart}). Due to demagnetization effects the magnetic field
profiles within the sample has peaks at the top
and bottom edges of the sample where the demagnetizing fields are minimal (Fig.~\ref{fig:SampleInPcell2}c).
On the other hand, the field profile
within the sample close to the center is quite homogeneous, since a
cylinder with infinite hight $H$ is equivalent to an ellipsoid in which the field is homogeneous.

In order to calculate the probability field distribution of a sample in a PC with a substantial
first moment  a model for the muon
stopping distribution is required. This distribution may be well approximated
by a three-dimensional Gaussian:\cite{SRIM}
\begin{equation}\label{eq:stopDistr}
P_s(x_1,x_2,x_3) = \frac{A}{(2\pi)^{3/2}} \prod_{i=1}^3 \frac{1}{\sigma_i}
\exp\left(-\frac{(x_i-x_{0,i})^2}{2\sigma_i^2} \right),
\end{equation}
where the subscripts $i=1, 2, 3$ correspond to  $x$, $y$, or $z$, respectively.
The quantities $x_{0,i}$ determine the mean value of the muon stopping
distribution, $\sigma_i$ are corresponding standard deviations, and $A$ is the normalization factor.
The quantities $x_{0,1}$, $x_{0,2}$, and $x_{0,3}$ can be determined quite accurately before starting
the experiment by tuning the momentum of the muon beam and vertical positioning of the sample.
For a sample with nearly the same density as the pressure cell $x_{0,1}\simeq x_{0,2}\simeq x_{0,3}\simeq 0$.
Simulations of the stopping distribution with the SRIM software \cite{SRIM}
yield $\sigma_1 = 0.875$~mm for copper (the basic component of the CuBe pressure cell) and
the minimal ratio of $\sigma_3/\sigma_1 = 3.36$. A maximal ratio of $\sigma_3/\sigma_1 \simeq 4$ is estimated
for the muon beam collimated by a $4\times 10$~mm collimator (this uncertainty is related with the degree
of muon beam focusing). The parameter $\sigma_2$ is in fact the standard
deviation of the function representing the convolution of a Gaussian with $\sigma = \sigma_3$ over the collimator
profile function along the $y$-axis. These parameters define the fraction of muons stopping in the PC
and the sample for a given sample geometry.
For a known $P_s(\mathbf{r})$ one can calculate the magnetic field probability
distribution $P(B)$ in the pressure cell by solving the integral:
\begin{equation}\label{eq:PB3D}
P(B) = \int\limits_{x^2+z^2>R^2} P_s(\mathbf{r})\delta(B-\mu_0[H + H'_z(\mathbf{r})]) d\mathbf{r}.
\end{equation}
Here, $\delta(x)$ is the delta function. The integration is taken over the volume of the pressure cell.
Note that this is not simply the probability field distribution in the pressure cell, but it is weighted with
the muon stopping probability distribution $P_s(x,y,z)$.
Fits of $P(B)$ to the experimental $\mu$SR data are shown in Fig~\ref{fig:Fig2PB}.
The function $P(B)$ describes the experimentally measured $\mu$SR signal rather well.

\end{document}